\documentstyle[12pt]{article}

\begin{document}

\input psfig

\centerline{\Large \bf PHASE TRANSITIONS IN THE UNIVERSE\footnote{
Invited article for {\it Contemporary Physics}.}}

\vspace{2.cm}

\centerline{Marcelo Gleiser\footnote{NSF Presidential Faculty Fellow.
email: gleiser@dartmouth.edu}}

\vspace{1.cm}

\centerline{\it Department of Physics and Astronomy}
\centerline{\it Dartmouth College}
\centerline{\it Hanover, NH 03755, USA\footnote{Permanent address.}}
\centerline{and}
\centerline{\it Nasa/Fermilab Astrophysics Center}
\centerline{\it Fermi National Accelerator Laboratory}
\centerline{\it Batavia, IL 60510, USA}
\centerline{and}
\centerline{\it Osservatorio Astronomico di Roma}
\centerline{\it Vialle del Parco Mellini, 84}
\centerline{\it Roma I-00136, Italy}

\vspace{2.cm}
\centerline{\large \bf ABSTRACT}
\begin{quote}
During the past two decades, cosmologists turned to particle physics in order to
explore the physics of the very early Universe. The main link between the
physics of the smallest and largest structures in the Universe is the
idea of spontaneous symmetry breaking, familiar from condensed matter physics. 
Implementing this mechanism into cosmology
leads to the interesting possibility that 
phase transitions related to the breaking of symmetries in high energy
particle physics took place during the early history of the Universe.
These cosmological phase transitions may help us understand many of the
challenges faced by the standard hot Big Bang model of cosmology, while
offering a unique window into the very early Universe and the physics of
high energy particle interactions.

\end{quote}
\vfill\eject

\noindent
{\large \bf I- The Universe we know}

\vspace{ 0.5cm}

\noindent
The fascination with grand questions is as old as time. Some of the earliest
records we have of ancient cultures, produced
long before what we now call science
existed, tell stories about the creation of the world or about the
Sun, the Moon, and the visible planets. Although
many empires have risen and fallen since then, and society has gone through
countless transformations, this fascination with grand questions has 
remained. It seems that
we just cannot avoid being curious about our origin, our future or that of the
world as a whole \cite{DANCE}.

With the development of modern science, 
physicists and astronomers have continued this
ancient tradition of asking grand questions about the world around us.
In 1609 Galileo
pointed a telescope to the skies for the first time, revealing a cosmos 
completely different from the then
prevalent Aristotelian view, while later in the
same century Newton unified the physics of earthly phenomena with that of
the skies through his law of universal gravitation.

But it was during the 20th century, through the marriage of Einstein's
new theory of gravity, the general theory of relativity,
and the construction of large telescopes, that 
physicists and astronomers
could really begin to face quantitatively
questions concerning the origin and future
of our world and of our place in it. Hence was born modern cosmology, the 
branch of physics which studies the properties and evolution of the
Universe as a whole.

In 1929, the American astronomer Edwin Hubble made the incredible discovery
that the Universe is expanding. Studying the spectra of galaxies, he
showed that the spectra were in general shifted towards the red, indicating
that these galaxies were moving away from us \cite{HUBBLE}.
Before him, a few theorists showed
that some of the solutions
obtained when applying Einstein's equations of general relativity to the
Universe as a whole implied that the Universe could be expanding
\cite{FRIEDMANN}. 

But what
does this really mean, an expanding Universe? First, a few words
about general relativity. According to Einstein's theory,
gravity can be understood as a deformation in the geometry of space and
in the flow of time due to the presence of matter. In the absence of matter,
space is flat and time flows undisturbed. One can think of a marble rolling
on a flat tabletop as a two dimensional analogy. When matter is present
things change dramatically. Imagine that the tabletop is made of an
elastic material. Now place a heavy lead ball in its centre. The table top
won't be flat anymore, due to the presence of the ball. This is a somewhat
simplified view of what happens to the geometry of space in the presence of
mass. The trajectories of the marbles will deviate from straight lines due
to the curvature of space. Thus, Einstein explained the acceleration caused
by gravity as being simply motion in curved space.

Now we can go back to the question of the expansion of the Universe. 
Since the main force controlling the expansion is gravity, the evolution
of the Universe will depend on its total mass. If the average density of
matter is equal or smaller than a critical value of about $10^{-29}{\rm
g/cm}^{3}$, the Universe will continue its expansion forever. Otherwise,
the Universe will collapse onto itself, possibly alternating cycles of
expansion and contraction. What is important to stress here is that the 
expansion of the Universe is an expansion of its geometry. Galaxies are 
moving away from each other because they are being carried by
the expanding geometry, somewhat like corks passively floating down a 
sloping river \cite{KT}.

In 1946, George Gamow, a Russian physicist working in the United States,
inspired by ideas from the Russian meteorologist-turned-cosmologist
Alexander Friedmann and from the Belgian physicist (and priest) George 
Lema\^{\i}tre,
proposed that the Universe emerged from a hot and dense soup of particles
and that it has been expanding ever since. Applying what was then known 
of nuclear physics to a mixture of thermodynamics and general relativity,
Gamow and his collaborators were able to reconstruct the history of the
first few minutes (actually about half an hour with nuclear data from 1950) 
of the Universe's 
infancy, starting from this primordial hot and dense soup of mainly
protons, neutrons, electrons, and photons. This model became later
known as the hot Big Bang model of cosmology \cite{GAMOW}.

The mathematical equations that dictate the evolution of the geometry in the
Big Bang model are quite simple \cite{KT}. 
Assuming that the Universe is homogeneous
(same everywhere) and isotropic (same in all directions), the geometry is
characterized by one single function of time, the scale factor $R(t)$ and
by one parameter $k$, which determines if the geometry is closed like that 
of a sphere ($k=1$), flat ($k=0$), or open ($k=-1$) like that of a saddle. 
Basically, the scale factor measures how the geometry stretches in time.
Assuming further 
that matter can be modelled by an ideal gas with energy density 
$\rho (t)$ and pressure $p(t)$, Einstein's
equations determine the dynamics of the scale factor $R(t)$ 
as follows,\footnote{
Unless otherwise specified,
the units here are chosen so that
$\hbar=c=k_B=1$, so that only mass or energy is relevant. It is customary
to measure energies in units of a GeV, i.e., $10^9$ eV, roughly
the mass of a proton divided by $c^2$. In these units, $G=m_{\rm Pl}^{-2}$,
where $m_{\rm Pl}= 1.2\times 10^{19}{\rm GeV} = 2.18\times 10^{-5}{\rm g}$ 
is the Planck mass.}

$$\left ({{\dot R}\over R}\right )^2 \equiv H^2 = 
{{8\pi G}\over 3}\rho +{k\over {R^2}}~,$$
and
$$2{{\ddot R}\over R}+\left ({{\dot R}\over R}\right )^2 = 
-8\pi Gp - {k\over {R^2}}~.$$
$H(t)$ is the Hubble factor. Note that $H^{-1}$ defines the time scale at which
quantities change in an expanding Universe. This time scale corresponds
to a length scale called the {\it Hubble radius}, $d_H(t)=H(t)^{-1}$. Processes
can only operate coherently within the Hubble radius.

The equations above are suplemented by the first law of thermodynamics,

$$d\left (\rho R^3\right )  = - pd\left (R^3\right )~.$$
These three equations are related by an identity called
the Bianchi identity. Thus, we can choose just two as independent equations.
The Friedmann models use the first and third equations.
For a simple equation of state, $p=\omega\rho$, we can find a relation
between the energy density $\rho$ and the scale factor $R(t)$.
There are three cases of interest, 
fixed by the choice of the constant parameter $\omega$:

$${\rm radiation}: \omega = {1\over 3} \Rightarrow \rho \propto R^{-4},$$

$${\rm cold~ matter}:\omega = 0 \Rightarrow \rho \propto R^{-3},$$

$${\rm vacuum~ energy}: \omega = -1 \Rightarrow \rho = {\rm constant}~.$$
Each of these cases is characterized by a different time evolution of the
scale factor: $R \propto t^{1/2}$ for radiation, $R\propto t^{2/3}$ for
cold matter, and $R\propto {\rm exp}(\sqrt{\Lambda_{\rm vac}}t)$ for vacuum 
energy,
where $\Lambda_{\rm vac}=8\pi G\rho/3$. The relevance of this last case will
become clear further on. In general, the total energy density is
a combination of all three kinds of matter. However, the evolution of
the scale factor is determined by the contribution that dominates the
energy density. For example, before $t_{\rm EQ}\sim 4\times 10^{10}{\rm sec}$,
radiation dominated over cold matter. Here we will be mostly interested in times
$t < t_{\rm EQ}$, that is, in a radiation-dominated Universe, the realm
of cosmological phase transitions.

What was remarkable about Gamow's proposal is that it made two crucial
predictions about our present Universe, which could be verified by 
observations. First, that the Universe should be permeated by electromagnetic
radiation with wavelength on the microwave region
and temperature of a few degrees above absolute zero. These were
the remnant photons from the epoch when hydrogen atoms were made,
roughly around 300,000 years after the bang (in today's numbers).

Second, that light nuclei such as deuterium (${\rm H}_1^2$), tritium (
${\rm H}_1^3$),
helium 4 (${\rm He}_2^4$), helium 3 (${\rm He}_2^3$), and 
lithium 7 (${\rm Li}_3^7$) were
cooked when the Universe was about 1 second old (in today's numbers), during
an epoch called ``nucleosynthesis'' and
with calculable proportions \cite{SCHRAMM}. Thus, Gamow and his 
collaborators predicted that
about $24\%$ of the Universe is made of ${\rm He}_2^4$.\footnote{The 
story
is somewhat more complicated. Initially, Gamow and his collaborators thought
that all elements could be synthesized in the early Universe. Only in the late
fifties it became clear than elements heavier than Li$_3^7$ were synthesized
in stars.}

During the past three decades, a growing amount of observational evidence
has offered very strong support for the Big Bang model. The cosmic background
radiation has been found by Penzias and Wilson in 1965, and is currently the
object of intense study by several groups in the world \cite{PENZIAS,CMB}. 
The abundances of
light nuclei have been observed to be consistent with the nucleosynthesis
predictions with one free paramenter, the relative excess of matter over
antimatter, something we will discuss further down. The fact that the Universe
was once very hot and dense as described by the Big Bang model is now 
widely accepted by the vast majority of physicists and astronomers.

\vfill\eject

\noindent
{\large \bf II - Challenges to the Big Bang model}

\vspace {0.5cm}
\noindent
But not everything is fireworks. As with any model in physics, the Big Bang
model has its limitations. It describes very successfully the evolution 
of the Universe from a hot and dense initial state, 
but fails to address several questions 
which are of great interest to modern cosmology. 
The application of atomic physics
to cosmology led to the understanding of the generation of the microwave
background radiation; the application of nuclear physics led to the
understanding of the formation of light nuclei during nucleosynthesis. Now,
in order to progress further with our understanding of the early Universe, 
we must
apply higher energy physics to cosmology. For the past two decades or so,
physicists have explored the possible consequences of applying
particle physics, the physics of subnuclear structures, to the Universe's
infancy. The topic of this article lies precisely in this interface between
particle physics and cosmology and on how this interface may help resolve
some of the challenges faced by the Big Bang model today. Before we move
on, here is a sample of questions that keep modern cosmologists busy.

\noindent
$\bullet$ {\it The smoothness problem}: 
One of the most startling properties of the
cosmic background radiation is its incredible smoothness. By smoothness 
I mean the lack of variarion in the temperature of the photons as measured in
all directions of the sky. When radio antennas are pointed at
different angles to measure the temperature of the cosmic 
background, they find it to be $T=2.726\pm 0.005 $ Kelvin, 
with deviations of less than 0.03\%.
These beautiful measurements present cosmologists with a serious problem.
Regions of the sky now separated by more than about a few degrees
were not in causal contact when the microwave radiation was produced,
as indicated in the figure. That is,
due to the finiteness of the speed of light,
interactions between particles that could insure thermal contact and 
homogeneity were not active beyond what today is a small patch in the sky. If
this is the case, how come all sky has the same temperature?

\begin{figure}
\hspace{1.3in}
\psfig{figure=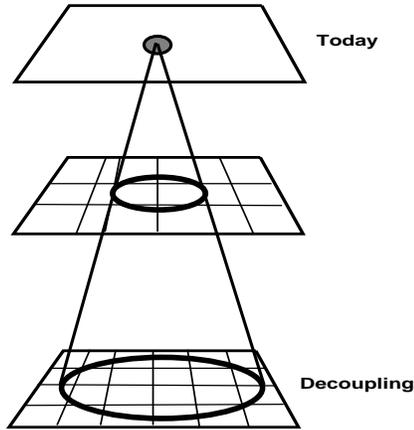,width=2.3in,height=2.3in}
\caption{The smoothness problem: A small patch in the sky today corresponds
to many causally disconnected regions during decoupling. Yet, the temperature of
the cosmic background radiation today is the same over the whole sky.}
\end{figure}

\noindent
$\bullet$ {\it The large-scale structure problem}: 
Some astronomers nowadays are
doing what great sailors of the 16th century did: extending the 
frontiers of the world, and mapping the new lands found beyond. Making a map of
the visible Universe is an extremely cumbersome task. Basically, one has to
identify each galaxy (actually, a statistically representative
sample of galaxies) and mark its
position in a three dimensional model of the cosmos. Apart from the obvious
impossibility of actually locating each galaxy (there are several billions
of them out there), one has to be able to measure its position, something which
is quite complicated when we are dealing with distances ranging from millions
to billions of light years away from us. As a consequence,
at this point in time, our maps
of the Universe are somewhat approximate, and may look to future generations
of astronomers as naive as the maps from the 16th century look to us today.

Still, our maps of the sky reveal something quite unexpected. 
In principle, we would expect the galaxies to be
scattered across the Universe without any sort of pattern, in a perfectly
random way. Instead, what is revealed by these maps is a richly-structured
Universe, where
galaxies tend to lay on vast sheet-like structures 
surrounding large empty regions,
somewhat like the foamy patterns we see in bubble baths \cite{GELLER}. 
Some of those
empty regions, or voids, have diameters of tens of Megaparsecs, or several
million light years. ($1{\rm Mpc}=3.26\times 10^6 
{\rm light~ years}= 3.09\times 10^{24}{\rm cm}$). Why would 
galaxies choose this rather particular 
distribution, as opposed to
being simply scattered across the sky without forming any obvious 
large-scale patterns?

\vspace{0.5cm}
\noindent
Figure 2: Voids: A three-dimensional view of voids in the SSRS2 survey. Voids
are regions (bubbles here) of space where very few galaxies are found.
[Figure can be found in H. El-Ad, T. Piran, and L. N. da Costa, Ap. J. Lett. 
{\bf 462}, (1996)
L13.]
\vspace{0.5cm}

\noindent
$\bullet$ {\it The matter-antimatter problem}:
Most of us learned in high school that matter is made of atoms and that
atoms are made of protons, neutrons and electrons. What we don't usually
learn in high school is that to each particle of matter there is another
particle, an ``anti-particle'', which is essentially the same as the
particle but with opposite electric charge. Thus, the electron has its
``anti-electron'', called a positron, which has positive electric charge,
the proton has an anti-proton, and so on. 
According to the laws of particle physics, matter and anti-matter should
be present in the Universe in equal amounts. And yet, we have ample
observational evidence that, at least in a very large volume 
extending far beyond our galaxy, there is much more matter than
anti-matter \cite{ANTIM,EWGLEISER}.

When particles collide with their anti-particles, the effects are devastating;
they both disintegrate into electromagnetic radiation, their energy
carried away
in photons.  In other words, if there were as
much matter as
anti-matter in the Universe, we wouldn't be here asking grand questions.
The Universe is somehow unbalanced, biased toward the existence of
matter over anti-matter. One of the greatest challenges in modern
cosmology is to unveil the roots of this cosmic imperfection.

Now that we have a sample of open challenges to cosmology, we can examine how
particle physics can help clarify them. As we will see, the application of
particle physics to early Universe cosmology will in many cases
invoke another branch
of physics, perhaps more familiar to everyday life than the physics of the
very large or the very small; the physics of phase
transitions.

\vspace{1.cm}
\noindent
{\bf III-Particles, Forces, and Symmetry Breaking}
\vspace{0.5cm}

Matter is organized in a hierarchical structure. Molecules are made of atoms,
and atoms of electrons orbiting nuclei made of protons and neutrons. 
Protons
and neutrons are examples of hundreds of particles found in
particle accelerators called {\it hadrons}, which (fortunately!) 
are not elementary, but made of yet smaller constituents called quarks. So far,
six quarks have been found. The distinctive feature of hadrons is that they 
interact via the strong nuclear force, the force responsible for keeping
the atomic nucleus together. They come in two types, {\it baryons} like
the proton and the neutron, which are made of three quarks, and {\it mesons},
which are made of a quark and an anti-quark. For example,
a proton is made of two {\it up} quarks and one {\it down} quark, while a
neutron is made of an {\it up} quark and two {\it down} quarks. 

According to modern particle physics, matter is made of two types of
elementary particles, quarks and leptons. The electron is a lepton, and so
is the muon. More recently another lepton has been found, the tau, which is
heavier than a proton. The name lepton, which comes from the Greek for
light weight, is a bit of an anachronism. Each of the three
leptons comes with its own neutrino, a massless and neutral particle. The
three neutrinos bring the number of leptons to six.

The distinctive feature of the leptons is that they interact via the other
force active at subnuclear distances, the weak force. In several
situations (but not all) when a lepton
interacts via the weak force, its associated neutrino appears. The best-known
example is beta decay, where a free neutron decays into a proton, an electron,
and its anti-neutrino with a half-life of approximately 10 minutes. 
[Or, in terms of quarks, $d\rightarrow 
u+e+{\bar \nu_e}$.]

The six quarks and the six leptons are the basic building blocks of matter.
They are neatly 
arranged in three families, as shown in the Table below. Only the
members of the first family make up matter familiar to us. Heavier quarks
and leptons appear as debris in very high energy collisons promoted by 
particle accelerators or some in cosmic rays. And, of course, in the hot 
furnace of the early Universe.

\vspace{0.5cm}
\hspace{1.cm}
\begin{tabular}{|c|c|c|c|}
\hline
 {\rm Types of Particles} & {\rm Family 1} & {\rm Family 2} & {\rm Family 3}  \\
\hline
 {\rm Leptons} & $e~~~\nu_e$ & $\mu~~~\nu_{\mu}$ & $\tau~~~\nu_{\tau}$  \\
\hline
 {\rm Quarks} & $u~~~d$ & $c~~~s$ & $b~~~t$  \\
\hline
\end{tabular}

\vspace{0.5cm}

\centerline{Table: Fundamental Building Blocks of Matter}

\vspace{0.5cm}

But this picture is not yet complete. In addition to identifying the basic
building blocks of matter, we must understand how these particles interact
with each other. Apart from the two short range forces mentioned above, the
strong and weak nuclear forces, there are, of course, two more
forces which, being long (actually infinite) range,
are very familar to us, the electromagnetic and the gravitational forces. These
four fundamental forces describe how the basic building blocks of matter
interact with each other. 

During the 19th century, mainly through the work of Michael
Faraday and James C. Maxwell, it 
became clear that the interactions between magnetic and electric bodies were 
best described in terms of the electromagnetic field. The concept of field
allowed for a local characterization of the interactions which was lacking in
the Newtonian notion of action at a distance. This notion of field has been
generalized to all four interactions between elementary particles. Furthermore,
following again the lead from the electromagnetic field and its quantization in
terms of photons, each field has its associated quantum (or quanta). 
The same is true for the particles themselves, considered 
quanta of their associated
matter fields. For example, an electron is a quantum of the ``electronic
field'', etc. Thus, we
arrive at a description of particles and their interactions in terms of 
interacting 
fields and their quanta. 

According to this description, there are two kinds of particles in Nature. The 
particles that make up matter (quarks and leptons), which are quanta of their
associated matter fields, 
and the particles responsible for their interactions, the quanta of the force
fields, namely
the photon (electromagnetism), the graviton (gravity), the eight gluons
(strong interaction), and the three vector bosons $W^+,~W^-,~{\rm and}~ 
Z^0$ (weak interaction) \cite{KANE}.

One of the great successes of particle physics is to have arrived at a
consistent mathematical description of how elementary particles interact with
each other up to energies of about 1000 GeV. 
This formulation is based on the so-called ``principle of gauge
invariance'' (PGI). In its simplest version, applied to electromagnetism, 
the PGI asserts that Maxwell's equations are invariant under certain 
transformations of the scalar and vector potentials. [Specifically,
$A_0({\bf x},t)\rightarrow A_0({\bf x},t) - (1/c) \partial \Lambda ({\bf x},t)/
\partial t$ and ${\vec A}({\bf x},t)\rightarrow {\vec A}({\bf x},t) +
{\vec \nabla} \Lambda ({\bf x},t)$, where $\Lambda ({\bf x},t)$ is an arbitrary
scalar function.]
The relativistic Hamiltonian
describing the interaction of a charged particle of mass $m$ and charge $e$
with an electromagnetic
field is also invariant under the same transformations ($c$ restored for
convenience),

$$ H(A_0,{\vec A}) = \sqrt{c^2 \left ({\vec p} - {e\over c}{\vec A}\right )^2
+m^2c^4} + eA_0 = H\left (A_0^{\prime},{\vec A}^{\prime}\right ) $$

The important point is that the interaction, or better,
the coupling, between the particle and
the electromagnetic field is uniquely fixed by the PGI. This is
seen through the terms coupling the particle's momentum to the electromagnetic
vector potential [${\vec p} - (e/ c){\vec A}$] and the term
$eA_0$, coupling the particle's charge to the scalar potential. Other
forms of coupling would not be ``gauge'' invariant.

The PGI is easily generalized for studying the dynamics of charged
fields as opposed
to charged particles. The coupling follows the same rule, but using the
substitution ${\vec p}\rightarrow -i\hbar {\vec \nabla}$,
familiar from quantum mechanics. This operator acts, for example, on a complex 
scalar field $\phi$ [that is, a field defined by two real functions,
[$\phi = (\Phi_1 \pm i\Phi_2)/{\sqrt 2}$] 
in its four dimensional relativistic 
generalization, $D_{\mu}\phi = (\partial/\partial t +ieA_0, {\vec \nabla}+
ie{\vec A})\phi$, where the index $\mu$ runs from 0 to 3. A complex scalar 
field represents electrically charged spin-0 particles.

A Lagrangian density (as we are dealing now with fields and not point
particles)
describing the coupling of scalar and electromagnetic
fields (also known as the Abelian-Higgs model)
is then built by squaring this operator with its complex conjugate
(the Lagrangian is a real function of the fields) and adding a kinematic term
for the electromagnetic field itself. To this, we could add a potential term 
for the field $\phi$, generally written as $V(|\phi|^2)$. Thus, the Lagrangian
density is written as \cite{RYDER}

$$ {\cal L}(|\phi^2|,A_{\mu}) = (D_{\mu}\phi)(D^{\mu}\phi)^{\ast}-
V(|\phi|^2) -[{\rm kinematic~ term~ for~} A_{\mu}], $$
where a summation is understood for the up and down indices. Note that the
interaction between the fields is built into the derivative terms. As in
ordinary Lagrangian mechanics, a variation of 
this Lagrangian density (henceforth Lagrangian)
with respect to the fields will generate their
equations of motion.

Now comes the beautiful part. 
As long as the scalar field $\phi$ transforms as $\phi \rightarrow 
e^{ie\Lambda}\phi$ (and its complex conjugate), 
where $\Lambda$ is the scalar function appearing in the
gauge transformation of the electromagnetic field, this Lagrangian 
is gauge 
invariant. In other words, making sure that the
Lagrangian is gauge invariant determines uniquely how the scalar field 
interacts with the electromagnetic field.

There is another crucial piece of information that comes
from the PGI. Above I mentioned that we should add a kinematic term to the
Lagrangian describing the dynamics of the charged-scalar and electromagnetic
fields, but said nothing of a mass term for the photon. ``Why should you?'',
the reader would ask, ``as we know the photon is massless anyway?'' Right, but
we can actually invert this statement and say that we know that the photon is
massless because a mass term for the photon would break the gauge invariance
of the Lagrangian! [The reader can easily verify this by adding a mass term
$m_{\gamma}^2A_{\mu}A^{\mu}$ for the photon in the Lagragian above. This
term does not remain invariant under the gauge transformation $A_{\mu}
\rightarrow A_{\mu} + \partial_{\mu}\Lambda ({\bf x},t)$.]

The point of this argument is that if we want to generalize the PGI to other
interactions, we must be careful. Let us consider the weak interactions. The
weak force carriers, the $W^{\pm}$ and the $Z^0$, known as the ``gauge 
bosons'',
are massive, as they should be for a
short range force. This being the case, how can we apply the PGI to the weak
interactions? Here is where one
of the most important ideas in modern particle physics comes to the rescue.
This idea is also 
the main link between particle physics and early Universe
cosmology. It is known as {\it spontaneous symmetry breaking}, and is 
inspired by similar ideas in condensed matter physics.

The gauge bosons are massive because at
low energies the weak interactions are not gauge invariant: The gauge symmetry
is broken at low energies.
However, at high energies the
gauge symmetry is restored and the gauge bosons are massless, just like the
photon. In other words, at
high energies the weak interactions become long range, just like
electromagnetism. Based on this idea, S. Glashow, A. Salam, S. Weinberg 
and others showed
that at sufficiently high energies the electromagnetic and weak interactions
can be unified into a single description, the {\it electroweak} theory. In fact,
their unified description predicted the existence of the gauge bosons, which
were observed in 1983 at CERN,
the European particle
physics laboratory located in Geneva, Switzerland. This very successful theory
is known as the Standard Model of particle physics.

The symmetry breaking is implemented by a Lagrangian similar to
the one we examined above. As in the classical mechanics of a massive
particle, a mass term
for the field appears as a quadratic term in the Lagrangian. For example,
the potential for a free massive scalar field is written as $V(|\phi|^2)= 
m^2|\phi|^2$. But in order to break the gauge invariance, it is the interaction
field (the photon in the Lagrangian above) that must get a mass.
It turns out that a mass term for the gauge boson is naturally
generated by the derivative terms in the Lagrangian. From the expression for
the derivative above,

$$(D_{\mu}\phi) (D^{\mu}\phi)^{\ast}= 
\partial_{\mu}\phi\partial^{\mu}\phi^{\ast}
+ e^2|\phi|^2A_{\mu}A^{\mu}+ {\rm other~ unimportant~ terms}.$$
We can see that the term proportional to $e^2$ can be interpreted
as an effective mass term
for the gauge boson, as long as the scalar field $\phi$ acquires a constant
value, known as the {\it vacuum expectation value} of $\phi$ or VEV, for
reasons that will become clear shortly. It is customary to denote the
VEV by $\langle \phi \rangle$.
This constant value of the field $\phi$ is fixed through its interactions,
that is, through its potential. In order to generate a mass term for the
gauge boson, the potential for the scalar field is written as

$$V(|\phi|^2)= m^2|\phi|^2 +\lambda |\phi|^4~~,$$
where $\lambda$ is the positive 
scalar self-coupling constant. There are two possible
cases, depending on the sign of $m^2$. If $m^2>0$ then the potential has only a
minimum at $|\phi|^2=0$, and the Lagrangian describes ordinary scalar
electrodynamics with 2 massive scalar particles with mass $m$ and a 
massless photon. Since this minimum is also the energy minimum for the
system (with all fields constant), this minimum is known as the ``vacuum'' of 
the theory and the value of $\phi$ there, the vacuum expectation value. 
But if $m^2<0$ the potential assumes the ``mexican hat''
shape, there is a maximum at the origin and a minimum at
$|\phi|^2 = -m^2/2\lambda \equiv \nu^2/2$ and  
the gauge boson gets a mass term. At this minimum, the gauge symmetry is broken.

In order to compute the correct masses, one must look at small fluctuations
about the minimum, or vacuum, of the theory. Leaving the details aside, one
finds that this theory describes a massive scalar field with mass
$m_{\phi}^2 =-2m^2$ and a massive vector field with mass $m_{A_{\mu}}=e\nu$.
Thus, the whole procedure can be summarized quite simply in
two steps: We first impose
that the Lagrangian describing the dynamics of the interacting
fields be gauge invariant.
This determines uniquely the couplings between the various fields. By choosing
the appropriate potential for the scalar field, 
the gauge invariance is spontaneously broken in the ``broken
vacuum'', where the gauge fields acquire a finite mass. 

\vspace{1.cm}
\noindent
{\bf IV-Phase Transitions}
\vspace{0.5cm}

How can we connect this whole description of spontaneous symmetry breaking
to cosmology? The answer is temperature. The main effect of including
temperature in the above description is to change the shape of the potential
$V(|\phi|^2)$. In particular, the main corrections to the potential come
in the mass term for the scalar field, which gains a positive contribution
proportional to the square of the temperature. Thus, including the leading
temperature corrections, the potential is written as,

$$V(|\phi|^2,T) = (-m^2 +\alpha T^2)|\phi^2|+\lambda |\phi|^4~~,$$
where $\alpha$ represents a combination of numerical factors and relevant
coupling constants. The crucial point to note here is that depending on the
value of the temperature, the coefficient of the quadratic term may be positive
or negative. For $T^2 > -m^2/\alpha$, the potential has only one minimum at
$|\phi|^2=0$, the theory is symmetric and
the gauge bosons are massless. For $T^2 < -m^2/\alpha$ the minimum is
away from the origin, the symmetry
is broken and the gauge bosons are massive. 
Thus, we can define the critical temperature $T_c^2 = -m^2/\alpha$,
below which the symmetry is broken. [The reader may also want to consult
the article by M. Shaposhnikov for a discussion of finite temperature
effects on symmetry breaking.] 

The early Universe was a very hot place. In order to describe the way particles
interacted we must include temperature effects. And since high temperatures 
restore symmetries, at sufficiently early times the electroweak symmetry
was restored! With parameters from the standard model of particle physics,
the electroweak symmetry was broken at about $10^{-12}$ sec after the bang,
when the temperatures were of the order of hundreds of GeV. For comparison,
the mass of the $Z^0$ is $m_{Z^0}=91~{\rm GeV}/c^2$. Thus, through its
history the Universe may have had different phases, where particle
interactions obeyed different symmetries.

There is another area of physics where the concept of symmetry breaking
is very familiar, {\it i.e.}, 
condensed matter physics, in particular in applications
of statistical mechanics to the study of phase transitions \cite{DOMAINS}. 
We all know of
phase transitions from everyday experiences as simple as boiling or freezing
water. More relevant to our discussion, we also know that temperature 
changes may affect
the symmetries of substances, just as it does the symmetries of particle
physics. As an illustration, let's consider water and its phases. In its liquid
phase, the probability that we will find a molecule of water somewhere in a 
container is fairly constant. As we drop the temperature and water freezes,
the molecules will arrange themselves in some kind of lattice and this 
uniformity will be lost. Thus, we can say that the decrease of temperature
decreased the symmetry of water. 

A more concrete example is that of a ferromagnet. There are two effects 
competing with each other, the interactions between nearby spins, which tend 
to align them, and the temperature, which tends to randomize their directions.
At high enough temperatures
the net magnetization of a sample is zero, as spins will have equal
probability of
pointing in all directions. This is the {\it paramagnetic} phase.
We say that the ferromagnet has a rotational
symmetry ${\cal O}(3)$, for the orthogonal group. The net magnetization can
be written, in a continuous approximation,
as ${\vec M}(T)=(1/V)\int {\vec s}({\vec x},T)dV$, where ${\vec s}({\vec x},T)$ 
is the local spin density.
As the temperature is lowered, temperature fluctuations will decrease and
nearby spins will tend to become aligned. As a result, the
ferromagnet will develop domains with a certain magnetization. If there is an
external magnetic field, all spins will tend to align with it. Thus, the 
original rotational
symmetry is broken below a certain temperature, the critical 
temperature, $T_c$.   

\setcounter{figure}{2}
\begin{figure}
\hspace{1.in}
\psfig{figure=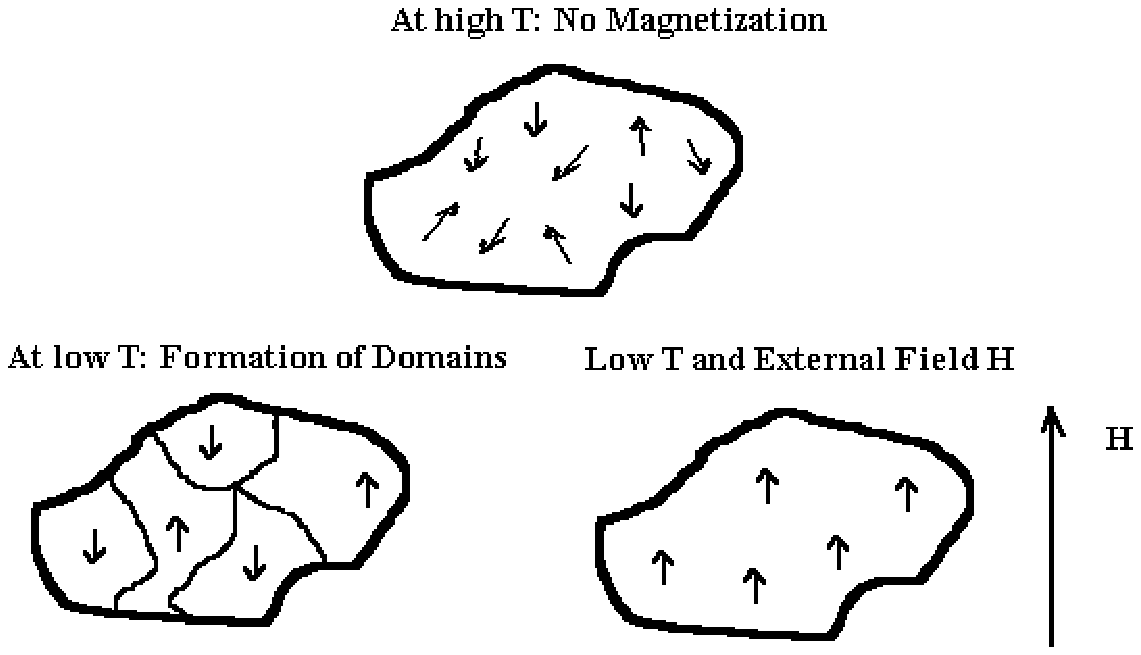,width=3.in,height=2.5in}
\caption{Domain formation during a continuous phase transition: At high
temperatures, there is no magnetization. At lower temperatures, individual 
magnets tend to align with their neighbours creating a local net 
magnetization, or magnetic domain. The interfaces of the domains move, until 
the whole sample has the same magnetization. An external magnetic field can 
bias the direction of the magnetization.}
\end{figure}

The connection between symmetry breaking in the early Universe
and phase transitions has led,
during the past 15 years or so, to
the emergence of a new interface in physics, namely, that of cosmology and
condensed matter physics. Thus, a cosmologist interested in the physics of
the early Universe will have to learn techniques from statistical mechanics
of phase transitions.
Terms
like spinodal decomposition and bubble nucleation are now part of the 
vocabulary of many cosmologists.

What is more important, if phase transitions indeed occurred in the early 
Universe they would have 
generated a host of possible observational consequences that
not only offer a window to very high energy physics but also may help us
solve some of the cosmological problems listed in Section II. 
Although I won't be
exhaustive here, I hope to give the reader at least a flavour of how this is
done in practice. In order to do so, 
I will concentrate on how ideas from phase transitions can be applied to
the resolution of the three
challenges to the Big Bang model explained above.

A beautiful and simple description of symmetry breaking in condensed matter
physics is presented by the Ginzburg-Landau theory \cite{LANDAU}. 
In order to describe
symmetry breaking, one must identify an {\it order parameter}, 
that is, a variable which describes the bulk properties of the system as 
a given {\it control parameter} changes. For example, 
for the ferromagnet the order parameter may be
the net magnetization and the control parameter may be the temperature, or, if
we have a mixture of two fluids, the order parameter may be the local
concentration difference of the two fluids, and the order parameter the 
temperature, etc. Order parameters may be scalar functions or complicated
matrices, as in the case of ${\rm He}^3$. For simplicity,
we will restrict ourselves to the
simplest possible example (already quite complicated!), that of a scalar
order parameter. These models describe liquid-gas transitions, binary
fluid mixtures, metal alloys, and Ising ferromagnets, ferromagnets where
the spins are restricted to point either up or down with respect to
some axis. All these models fall in the so-called {\it Ising universality
class}, that is, have critical properties which are essentially identical.

For a scalar order parameter, the homogeneous part of the Ginzburg-Landau
free energy density is written as

$$V_{\rm GL}(\phi,T)= {1\over 2}\alpha(T)\phi^2 + {{\lambda}\over 4}
    \phi^4 -{\cal H}\phi~~,$$
where $\alpha(T)\equiv a(T-T_c)/T_c$ and ${\cal H}$ 
is an external magnetic field. The similarity with the finite temperature
potential for the scalar field is striking! Again, depending on the value of
the temperature, $\alpha(T)$ can be positive or negative, being zero at $T_c$. 
The values of the order parameter $\phi$ above and below $T_c$ will depend on 
the external magnetic field ${\cal H}$, as shown in the Figure.

\begin{figure}
\begin{minipage}{.46\linewidth}
\psfig{figure=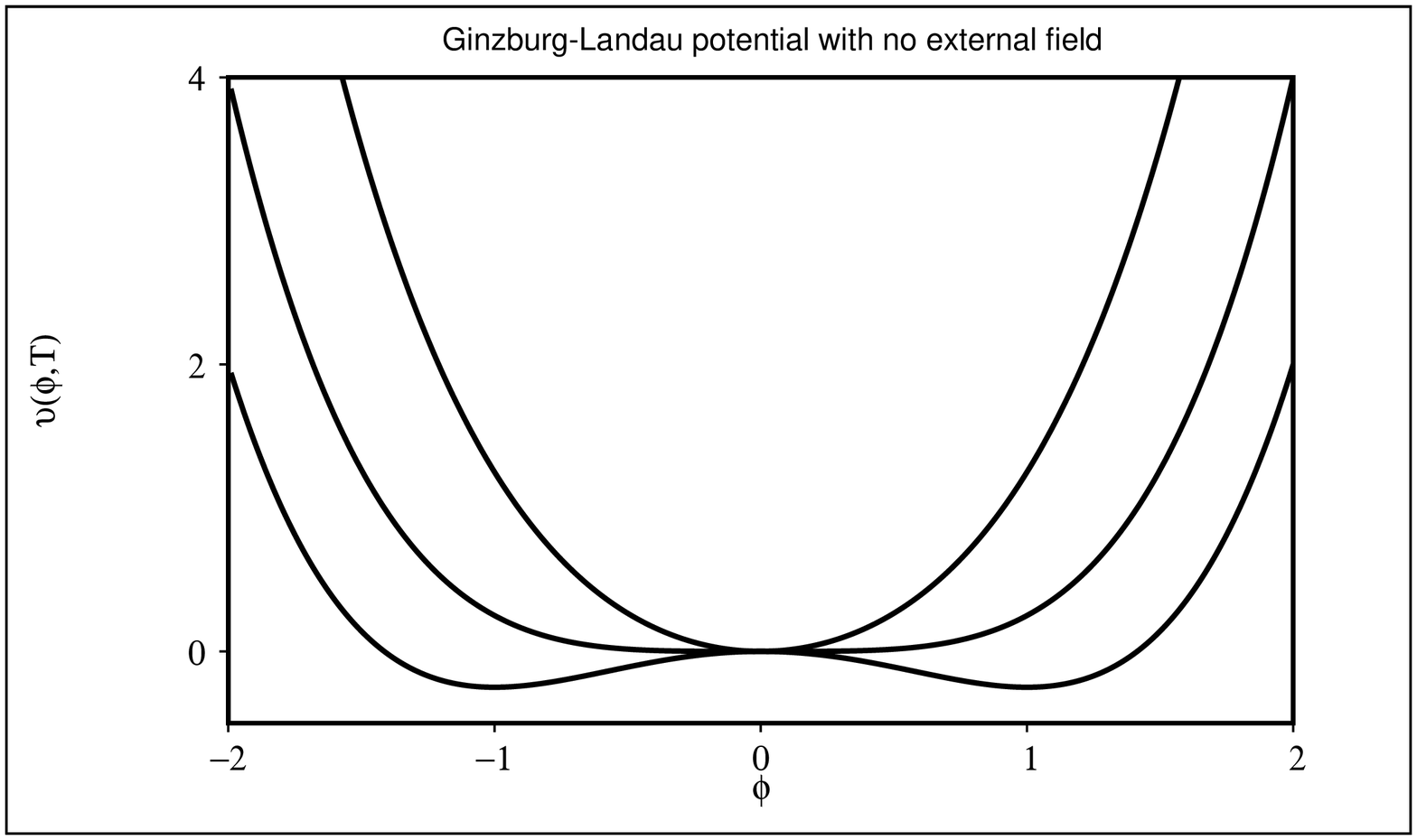,width=\linewidth,height=2.in}
\end{minipage}
\begin{minipage}{.46\linewidth}
\psfig{figure=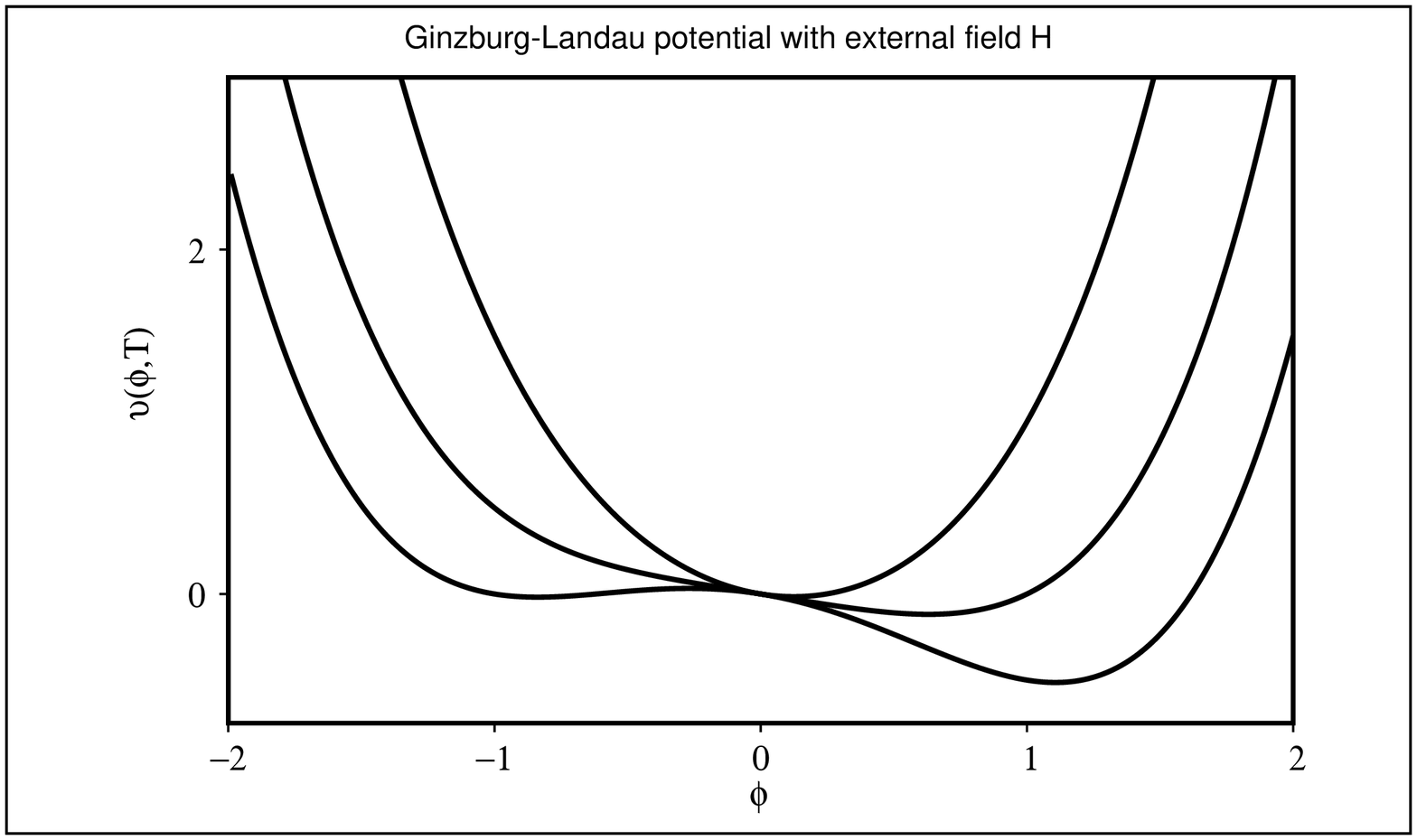,width=\linewidth,height=2.in}
\end{minipage}
\caption{Ginzburg-Landau potentials: Different Ginzburg-Landau potentials
for a scalar order parameter are shown without (left) and with (right) an
external magnetic field.}
\end{figure}

The reader can see that there are basically two possible kinds of phase
transitions depending on the value of the external field ${\cal H}$. 
If ${\cal H}=0$, the
free energy is symmetric with respect to a $\phi \rightarrow -\phi$ 
reflection and is simply a degenerate double well. As the temperature drops
below $T_c$, the order parameter changes {\it continuously} between the
symmetric $\phi=0$ and the broken-symmetric phase with $\phi^2 = 
-\alpha/\lambda$. This is an example of a {\it continuous} phase transition,
sometimes called a second order phase transition based on an old classification
by P. Ehrenfest. The heat capacity $C_V= -T\partial^2V_{\rm GL}/\partial T^2$
has a discontinuity at $T_c$. 

Continuous phase transitions evolve by {\it spinodal decomposition} 
\cite{LANGER}. Basically,
long-wavelength small-amplitude fluctuations grow exponentially fast as
domains of the two phases form and compete for dominance. These domains are
separated by an interface, where the order parameter changes continuously
from one phase to another. Clearly, the order parameter only provides 
information of the global behavior of the system, leaving aside the
complicated local dynamics of the interfaces, a topic of much interest for
researchers.

\vspace{0.5cm}
\noindent
Figure 5: Spinodal Decomposition: The system is initially prepared in
thermal equilibrium at $< \phi > =0$. It is then suddently cooled
and left to relax to its lowest free energy state. The formation of an
interface can be easily seen. [The color PostScript file can be obtained
by request from Carmen Gagne at: carmen.gagne@dartmouth.edu]
\vspace{0.5cm}

If the external field $H$ is not zero, the free-energy density will not be
degenerate anymore. The presence of the field will bias the system, determining
the lowest free energy phase. If the system starts in the highest free energy
phase at high temperatures and is rapidly cooled (or quenched)
to below $T_c$, it will
remain trapped in the high free energy phase. In this case, we say that the
system is in a metastable phase. The transition to the lowest free energy
phase is {\it discontinuous}, and occurs through the nucleation of bubbles
of the low free energy phase within the metastable phase. This is an
example of a discontinuous or first order phase transition.
The phenomenon of bubble nucleation is a beautiful illustration
of nonlinearities in action;
the bubble configuration represents a coherent fluctuation of the field,
excited by thermal or quantum effects. We will say more about quantum 
nucleation below.
Let us examine this mechanism
in a little more detail. 

\setcounter{figure}{5}
\begin{figure}
\hspace{1.in}
\psfig{figure=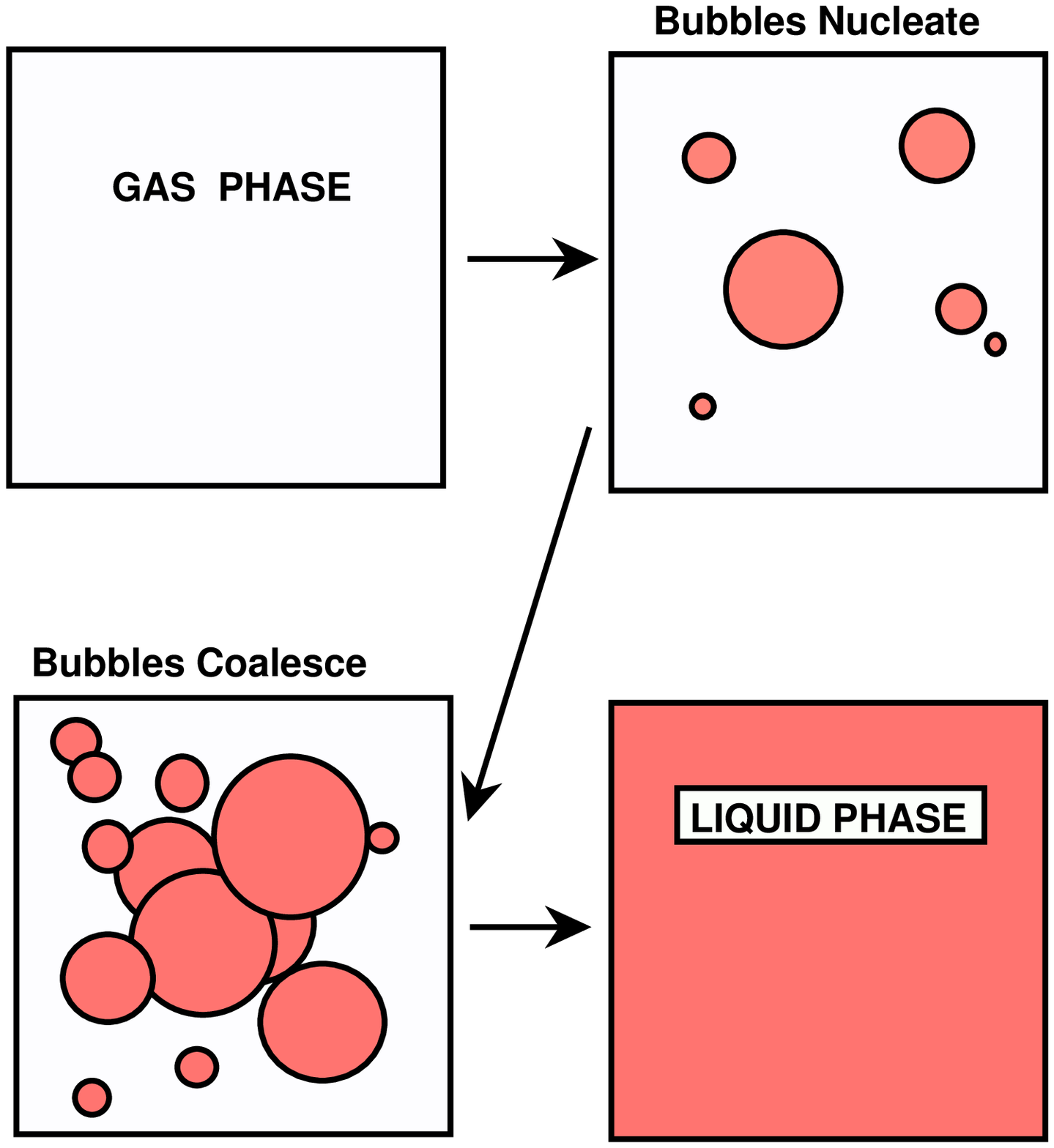,width=2.5in,height=2.5in}
\caption{Bubble Nucleation in a discontinuous phase transition: Bubbles
from the liquid phase appear inside the gas phase. If they are large enough
they grow and percolate, eventually converting the whole volume into the
liquid phase.}
\end{figure}

The energy of a field configuration $\phi_c$ is given by

$$E[\phi_c] = \int dV \left[ {1\over 2}{\vec \nabla}\phi_c\cdot 
{\vec \nabla}\phi_c + V_{\rm GL}(\phi_c)\right ]~~. $$
It is convenient to analyse the expression for the energy
of a {\it thin-wall} bubble, that is, a spherically symmetric 
configuration with a well-defined interior of radius $R$, separated
from the exterior by a wall of thickness $L \ll R$.
Thus, we can divide the 
configuration into three parts: the ``inside'', where $\phi_c={\rm const}=
\phi_+$; the ``outside'', the metastable phase, where $\phi_c={\rm const}=
\phi_-$; and the bubble wall, where the field interpolates between the
two minima, $\phi_{\rm w}(x/L)\propto \tanh(x/L)$. Within this approximation
we can write, for the energy of a thermally nucleated bubble,

$$E[R] = 4\pi \sigma R^2 - {{4\pi}\over 3}\epsilon R^3~~,$$
where $\sigma \equiv  (1/2)\int dr {\vec \nabla}\phi_c\cdot{\vec \nabla}\phi_c$
is the surface density and $\epsilon \equiv |V_{\rm GL}(\phi_+,T) -
V_{\rm GL}(\phi_-,T)|$ is the free-energy difference between the two phases.
Thus, there is a critical radius $R_c = 2\sigma/\epsilon$ above which it is
favourable for the configuration, or bubble, to grow. If $R<R_c$, the
surface tension dominates and the bubble shrinks. Note that if
$\epsilon = 0$, that is, if the two phases are degenerate, $R_c\rightarrow
\infty$. $R_c$ also determines
the free energy barrier for the nucleation of the critical bubble,
$E[R_c] = (16\pi/3)\sigma^3/\epsilon^2$. The nucleation rate per unit time and
unit volume is approximately written as

$$\Gamma \propto T^4 \exp [-E(R_c)/T]~. $$
If bubbles with $R>R_c$ start being nucleated they will grow and coalesce,
eventually converting the whole volume of the system from the metastable phase
to the lower free energy phase, completing the phase transition. (See Figure
6.)

A similar bubble nucleation mechanism is possible via quantum fluctuations
as opposed to thermal fluctuations. The field, trapped in the metastable
state, will {\it tunnel} to the ground state with a given rate calculated in
a similar way. This tunneling is the field theoretical equivalent of
the usual barrier penetration mechanism in nonrelativistic quantum
mecyhanics, where the wave function has a nonzero probability flow
through a potential barrier. As one would expect, 
the ``vacuum decay'' rate is smaller than the
thermal decay rate, $\Gamma_{\rm vac}\propto m^4\exp [-S(R_c)/\hbar]$,
where $S(R_c)\propto \sigma^4/\epsilon^3$ and $m$ is the typical mass
scale in the problem. $S(R_c)$ is the four dimensional
Euclidean action for the critical bubble configuration, a measure of the 
barrier for quantum tunneling.
Both thermal and quantum bubble
nucleation may play a role in the early Universe.
Next we will examine how these ideas are applied in the context of early 
Universe cosmology.

\vspace{1.cm}
\noindent
{\large \bf V-Cosmological Phase Transitions}
\vspace{0.5cm}

\noindent
In order to gain some insight into what effects the expansion of the Universe
may have in the dynamics of a phase transition, we must first remember that
the expansion rate determines how
fast the temperature drops compared to the interaction time scale of
the particles. From Friedmann's equations above, the expansion rate of
the Universe for a radiation-dominated era is (we can safely take 
the curvature constant $k=0$
in this regime)

$$ {{\dot R}\over R}= H\simeq 1.66g_{\ast}^{1/2}{{T^2}\over 
{ m_{{\rm Pl} } } }~~,$$
where I used that for a relativistic gas in thermal equilibrium at 
temperature $T$, $\rho = {{\pi^2}\over {30}}
g_{\ast}T^4$, where $g_{\ast}$ is the number of relativistic
degrees of freedom. A given
particle species is in thermal equilibrium in an expanding Universe as long as
its interaction rate is faster than the expansion rate, $\Gamma_{\rm int}> H$.
With mild assumptions about how particles interact at very high energies, this
happens as long as $k_BT\leq 10^{16-17}$ GeV \cite{KT}. 

So far I have motivated the discussion of cosmological phase transitions
using only 
the electroweak phase transition as an example.
Now, I would like to discuss another phase transition which
may have ocurred much earlier than the electroweak symmetry breaking.
This is a phase transition related to the breaking of the symmetry described
by the so-called Grand Unified Theory (GUT), 
where the strong interaction is unified with the
electroweak interaction \cite{ROSS}. Current estimates of when this unification
occurs lead to an energy scale of about $10^{16}$ GeV, roughly 14 orders
of magnitude above the electroweak unification. In the context of the
Big Bang model, such energies were achieved when the Universe was about
$10^{-38}$ sec old. 

If a phase transition indeed happened at this early time,
it would have left some very interesting signatures. One of them, 
which will be treated in future
articles by A. Gill and T. Vachaspati in this magazine, is the possibility
that the transition generated certain ``energy knots'' in the field
configurations, known as {\it topological defects}. The various types
of topological defects are determined by the kind of symmetry that is broken,
which in turn depends on the details of the unification scheme. For example,
if a discrete (left-right) symmetry is broken as in the simple GL model
above, sheet-like defects known as ``domain walls'' or interfaces
would appear between the two phases \cite{GUTPT}. 

These topological defects are both a blessing and a curse for cosmologists.
Curse because some of them may dominate the energy density of the Universe
and cause an expansion rate different from the observed one. 
Domain walls would create an ultra fast expansion rate, where the scale factor
would expand as $R\propto t^2$ \cite{GGK}. This expansion rate would lead
to severe discrepancies with observations, for example the abundance of
light elements predicted from nucleosynthesis.
Thus, cosmology rules out this kind of
symmetry breaking at the GUT scale, a beautiful example of how it
can influence particle physics. (Unless, of course, the left-right symmetry
is not exact...) Blessing because these topological defects may play
a role in the formation of large-scale structure, although models are becoming
increasingly constrained by observations of the cosmic background radiation.

But here I want to explore another possible
consequence of the GUT-scale phase transition, known as the {\it inflationary
cosmology}, as it exemplifies very clearly the interplay between cosmology
and the dynamics of phase transitions. Since the original model proposed by
the American physicist Alan Guth in 1981, the idea of inflation has mutated
into several different scenarios, some of them not invoking directly a 
GUT-scale symmetry breaking \cite{INFLATION}. However, in this author's opinion,
these various scenarios are viability studies which 
indicate how a larger theory 
related to unification at high energy scales will be applied to cosmology. 
With
this said, I move on to explain the basic mechanism of inflation, starting
with Guth's original model.

In Guth's model, the effective
potential responsible for the breaking of the GUT symmetry had a metastable
phase. As the Universe expanded and cooled, the scalar field responsible for
the symmetry breaking became trapped in this metastable state. In this case,
at high enough temperatures, the energy density of all matter would have two 
terms, one from relativistic radiation and one from the vacuum energy,

$$\rho = {{\pi^2}\over {30}}g_{\ast}T^4 + V_0~~, $$
where $V_0$ is the constant energy density of the field while 
trapped in the metastable
minimum. Clearly, when the temperature 
$T < \left ( 30 V_0/g_{\ast}\pi^2\right )^{1/4}$, the energy density will be 
dominated by the constant vacuum energy, and, as we saw in Section 1,
the scale factor will expand
exponentially fast, $R\propto {\rm exp}(\sqrt{\Lambda_{\rm vac}}t)$. 
Hence the name inflation. While the scalar field is trapped
in this metastable phase, the Universe will expand superluminally and
become supercooled. [Note
that this doesn't mean that information will be travelling faster than the
speed of light. Particles, and their interactions, will still obey
causality.] Due to the sudden drop in temperature, quantum effects will 
dominate over thermal effects.
Eventually, the field will decay into the lowest energy
state by the mechanism of quantum bubble nucleation. These bubbles will grow
and coalesce, completing the phase transition.

The problem with this original scenario, now called ``old inflation'', is
that the bubbles's growth rate, which is determined by causal processes,
can not compete with the exponential expansion rate of
the Universe; the bubbles can not meet and coalesce to complete the
phase transition.
Many alternatives have been proposed since then,
with names like new inflation, chaotic inflation, extended inflation,
natural inflation, etc. In one way or another, all scenarios for inflation
rely on the extra potential energy available as
a given scalar field (or fields)
approaches its final low energy state, be
it through bubble nucleation, or by rolling down its potential.
At this point it is fair to say that we understand
what are the desirable and undesirable
consequences of an inflationary epoch in the early
Universe, although we are still lacking a truly compelling model. Information
coming from GUT-scale particle physics and from the cosmic microwave
background would be most welcome here.

What are the benefits of an inflationary phase? First, if there are any
unwanted relics from the phase transition, such as undesirable topological
defects,
these would be ``inflated away'', that is, their number density would decrease
to a negligible and harmless amount. Second, for a sufficiently long
inflationary expansion, a small causally connected volume could have
grown to encompass the whole of the observed Universe today. In this case,
the smoothness problem of the Big-Bang model could be solved (see Figure 7).

\begin{figure}
\hspace{3.cm}
\psfig{figure=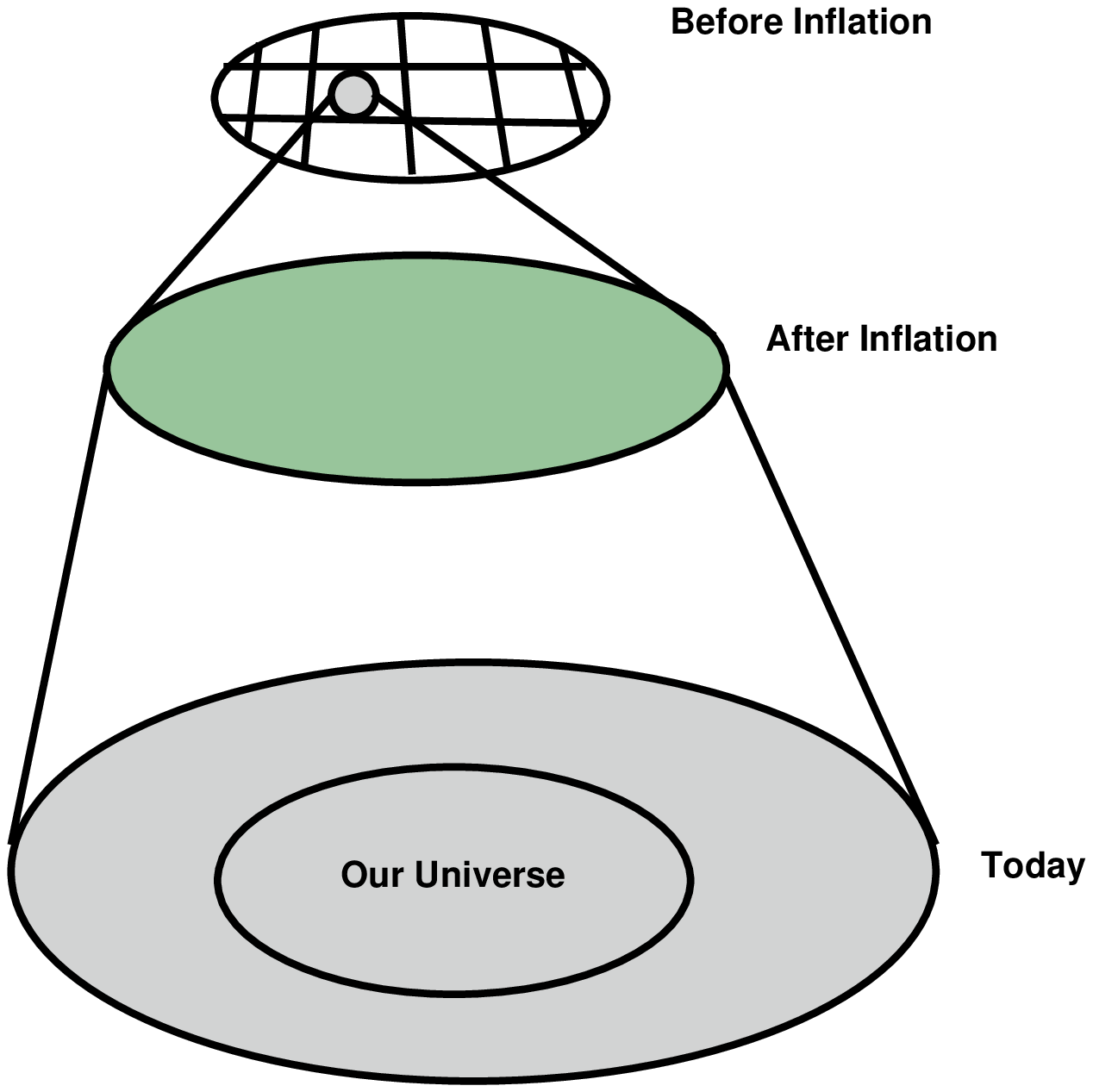,width=2.3in,height=2.3in}
\caption{Smoothness problem solved by inflation: Due to the exponentially
fast expansion during inflation, a small, causally-connected patch (shown
in grey) can grow to encompass our whole visible Universe.}
\end{figure}

In order to solve the smoothness problem, 
the scale factor must grow by about 30 orders of magnitude during
inflation, such that a length scale which before inflation was 
$10^{-28}{\rm cm}$ would
be ``stretched''
to $10^{2}{\rm cm}$ by the end of inflation. Today, this
length scale would correspond to $10^{28}{\rm cm}$, roughly the size of the
observable Universe. The important point is that before inflation, 
length scales smaller than $10^{-24}{\rm cm}$ were causally connected. Thus,
our whole observable Universe would have fitted quite comfortably within a
causally connected patch!

A second benefit from the inflationary model is that it can also provide a 
mechanism to generate the seeds that will be ultimately responsible for
the observed large-scale structure of the Universe. 
As we know from basic quantum mechanics, the zero-point energy of a quantum
system indicates the presence of fluctuations about the classical 
minimum of energy. An
example is the simple harmonic oscillator in the position representation. As we
move on to fields, the vacuum will also be populated by zero point
fluctuations of different wavelengths. Since during inflation length scales
are stretched exponentially fast ($R\propto {\rm exp}(\sqrt{\Lambda_{\rm vac}}
t)$,
while the Hubble radius ($D_H(t)=H(t)^{-1} = R/{\dot R}$) remains
constant, it is possible for a perturbation of a
given length scale to grow bigger than the Hubble radius
during inflation and then
reenter it at a later time during the radiation or the matter dominated eras
(see Figure 8)\footnote{Recall that a given length scale 
will grow at $t^{1/2}$ or $t^{2/3}$ during the radiation and matter
dominated eras, respectively, while the Hubble radius will grow as $t$.}. 
Thus, inflation offers a
mechanism of amplification of quantum fluctuations into the density
perturbations which will cause the gravitational instabilities needed
for structure formation later on in the evolution of the Universe.

\begin{figure}
\hspace{1.in}
\psfig{figure=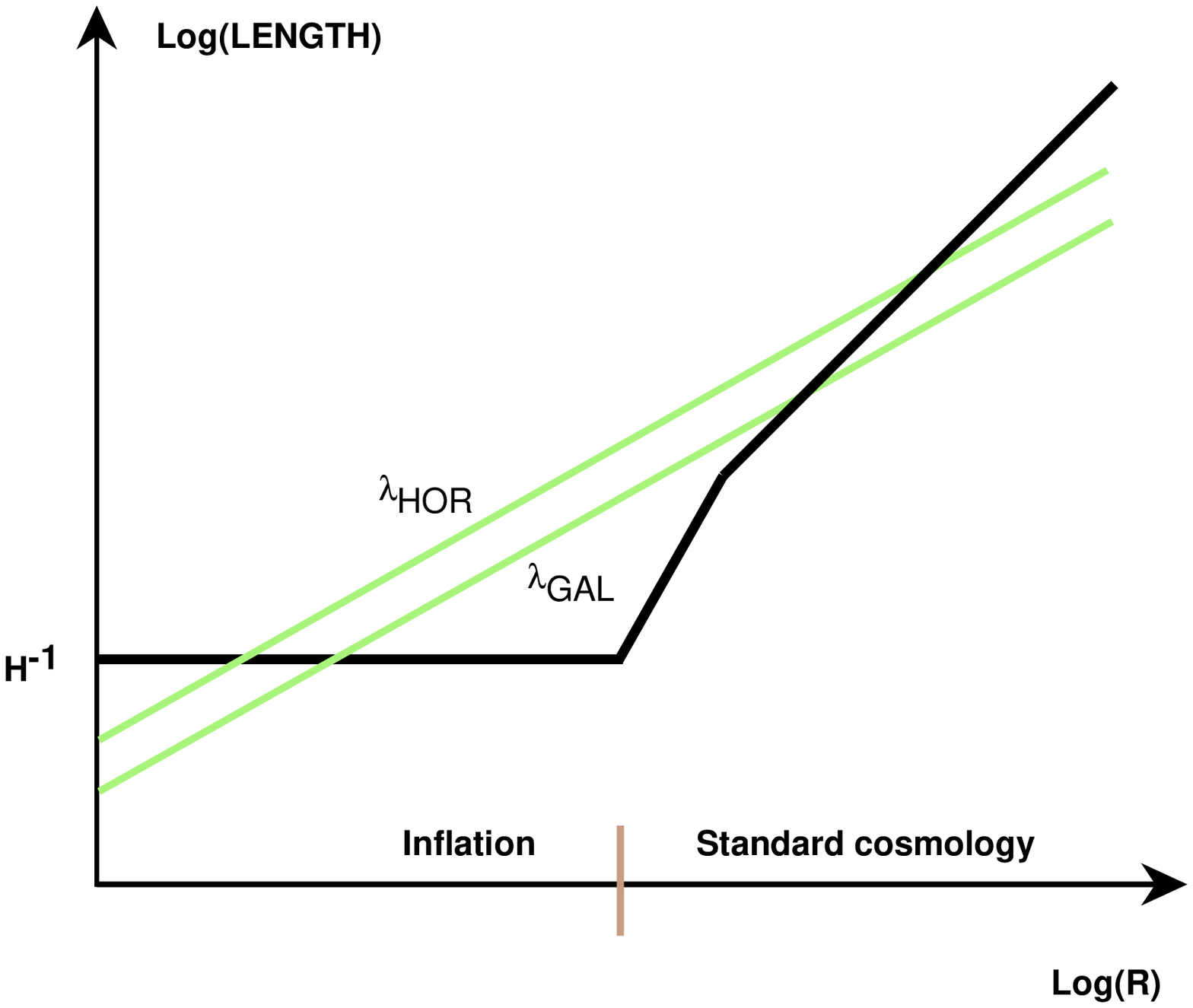,width=2.3in,height=2.3in}
\caption{Perturbations reentering causally connected horizon: During 
inflation, perturbations of both galactic ($\lambda_{\rm GAL}$) and horizon 
scale ($\lambda_{\rm HOR}$) are stretched outside causally-connected regions 
defined by the Hubble radius ($H^{-1}$). As inflation ends, the Universe
enters a radiation-dominated era, followed by a matter-dominated era, 
grouped jointly in the figure as ``standard cosmology''. During the standard
cosmology era, the perturbations grow slower than the Hubble radius and will
eventually reenter the
causally-connected Universe, generating the perturbations responsible for
large-scale structures of different sizes.}
\end{figure}

Finally, we will
examine how the third of the challenges to the Big Bang model we
mentioned earlier, that of
the excess of matter over anti-matter, can be solved by a primordial phase
transition. There are several mechanisms for generating the matter
(or baryonic) excess, either during the GUT phase transition or the
electroweak phase transition. However, here 
we will focus more on {\it electroweak
baryogenesis} as it calls for physics of much lower energy scales.
The basic ideas, applicable to both situations,
were presented in a pioneer
work by A. Sakharov in 1968. [The reader interested in more details should 
consult the article by M. Shaposhnikov dedicated to electroweak baryogenesis.] 

Sakharov 
suggested that three conditions must be satisfied in order to produce the
matter excess; first, there must be a way of creating both more baryons and
anti-baryons.
Then, there must be a
mechanism to bias the creation of more baryons than anti-baryons. And finally,
once we have an excess of matter particles over their anti-matter partners,
we must make sure that this excess is not erased as the Universe continues to
expand \cite{ANTIM,EWGLEISER}.

The first of these conditions is the creation of both baryons and
anti-baryons from collisions involving the other particles
present in the primordial soup.
At low energies, the number of baryons participating in
collisions between different particles is conserved, that is,  just like
electric charge, the total number of baryons before an interaction
equals the total after. If we are interested in making baryons, as we
must in order to create matter in the Universe, this conservation law is
not very useful. According to Sakharov's requirement, however,  at very
high energies the interactions between elementary particles should not
conserve the number of baryons. This is both true from the decay of heavy
GUT-scale particles and from the non-trivial vacuum structure of the
electroweak theory, which has degenerate minima with different baryon
number, as shown in Figure 9.

But this first condition does not differentiate between baryons and
anti-baryons. At high temperatures we could still create the same number of
each, and that wouldn't cause a bias toward matter over anti-matter. We
need a second condition. Once the high energies of the early Universe
allow for the creation of baryons and anti-baryons, we need a condition
that selects, or biases, the creation of baryons over anti-baryons, an
arrow pointing in the right direction ({\it i.e.}, toward baryons). This
is known as CP violation, from the operations of charge conjugation and
parity, familiar from quantum mechanics. In 1964, J.H.
Christenson and collaborators found experimental evidence that interactions
between certain baryons do indeed exhibit this bias. It is as if
Nature has its own biases, in this case toward more baryons. If this is true in
laboratory experiments, no doubt this will also be true in the early
Universe.

Finally, once we have produced a net baryonic excess, we must ensure that
it will not go away. This leads to the third Sakharov condition, that
baryogenesis requires the Universe to be out of thermal equilibrium.
We can understand the need for this condition with the following illustration.

Hot systems have no memory of their past.
Imagine a coffee spoon which is initially cold. Now immerse
one of its ends into a very hot cup of coffee. What happens? Although
early on only the end in the coffee will be hot, very quickly the whole
spoon will be equally hot. You won't be able to tell which of the two
ends was immersed into the coffee cup; the system (coffee spoon and hot
coffee) lost its ``memory''. Another term for this loss of memory is
thermal equilibrium. If the early Universe was in thermal equilibrium,
any excess baryons would have been deleted; in equilibrium, the net baryon
number is zero. In order to maintain the baryon bias as the Universe cools,
we need to make sure the Universe doesn't ``lose it's memory'' and delete the
new baryons. We need out of equilibrium conditions.

In GUT-scale baryogenesis out of equilibrium conditions are achieved by the
irreversible decays of heavy particles. Due to the expansion of the Universe,
the reaction rate of processes 
involving the decays of a heavy particle into lighter ones
won't be able to keep up with the expansion rate of the Universe. Roughly,
the expansion makes it hard for the lighter particles to meet and keep the
reaction 
going both ways. [$X \rightarrow q_i + {\bar q_j}$, where $q_i$ and
$q_j$ are two lighter particles.]

In electroweak baryogenesis, the idea is to use the dynamics of the phase
transition to generate the excess baryon number. It is clear that a phase
transition always involve out of equilibrium conditions: whenever the system
finds itself in a higher free energy density phase, it will relax into the
lowest possible free energy phase. Most mechanisms of electroweak
baryogenesis assume a typical discontinuous phase transition via the usual
bubble nucleation mechanism. 

In the symmetric high temperature phase, baryon number is freely violated
with a rate proportional to $T^4$. However, in the broken-symmetric, 
low-temperature phase, this rate is suppresed by a Boltzmann factor,
$\Gamma_{\rm b-viol}\propto \exp{[-E_{\rm sph}/T]}$, where $E_{\rm sph}$ is
the energy of the so-called {\it sphaleron} configuration (from the Greek
``ready to fall''), which interpolates
between vacua of different baryon number (see Figure 9). Thus, the excess
baryons are generated in the symmetric phase.

\begin{figure}
\hspace{1.in}
\psfig{figure=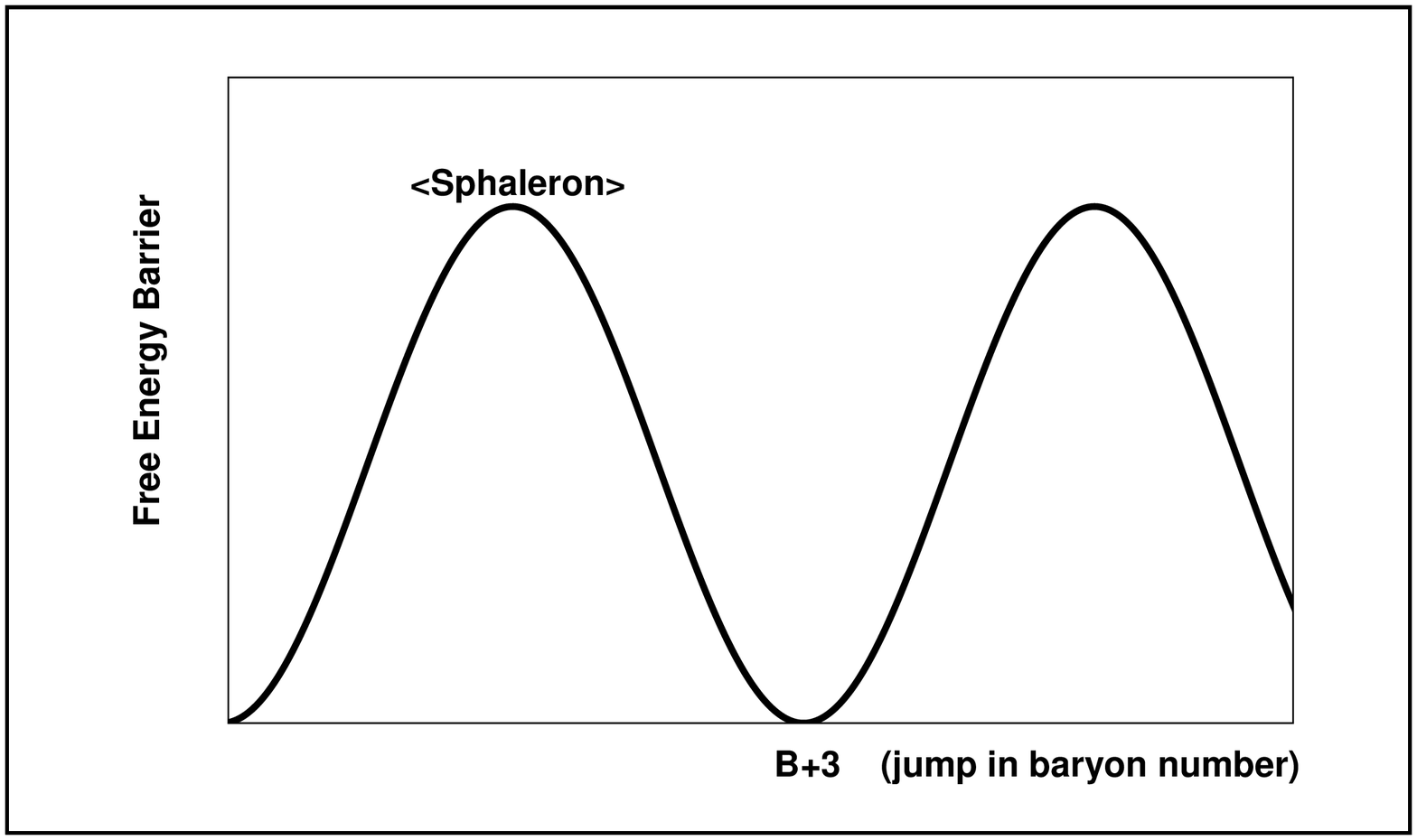,width=2.3in,height=2.in}
\caption{Baryon number violation in the electroweak theory: The nontrivial
vacuum of the electroweak theory allows for degenerate minima with different
baryon number. Thus, jumping from one vacuum to another it is possible to
generate the baryon number excess. The field configuration responsible for
the interpolation between different vacua at high enough temperatures is the
sphaleron. At very high temperatures, higher than the barrier shown, the
baryon number violation rate is simply proportional to $T^4$.}
\end{figure}

As the Universe expands and cools, bubbles of the broken-symmetric phase form
inside the symmetric phase. The excess baryons ``cooked'' in the symmetric
phase have a probability of going through the bubble wall, generating a
net baryon number excess inside the bubble. Since inside the growing bubbles
baryon number
is conserved (up to exponential accuracy), this net excess survives and
becomes the matter we are ultimately made of. (See Figure 10.)

\begin{figure}
\hspace{1.in}
\psfig{figure=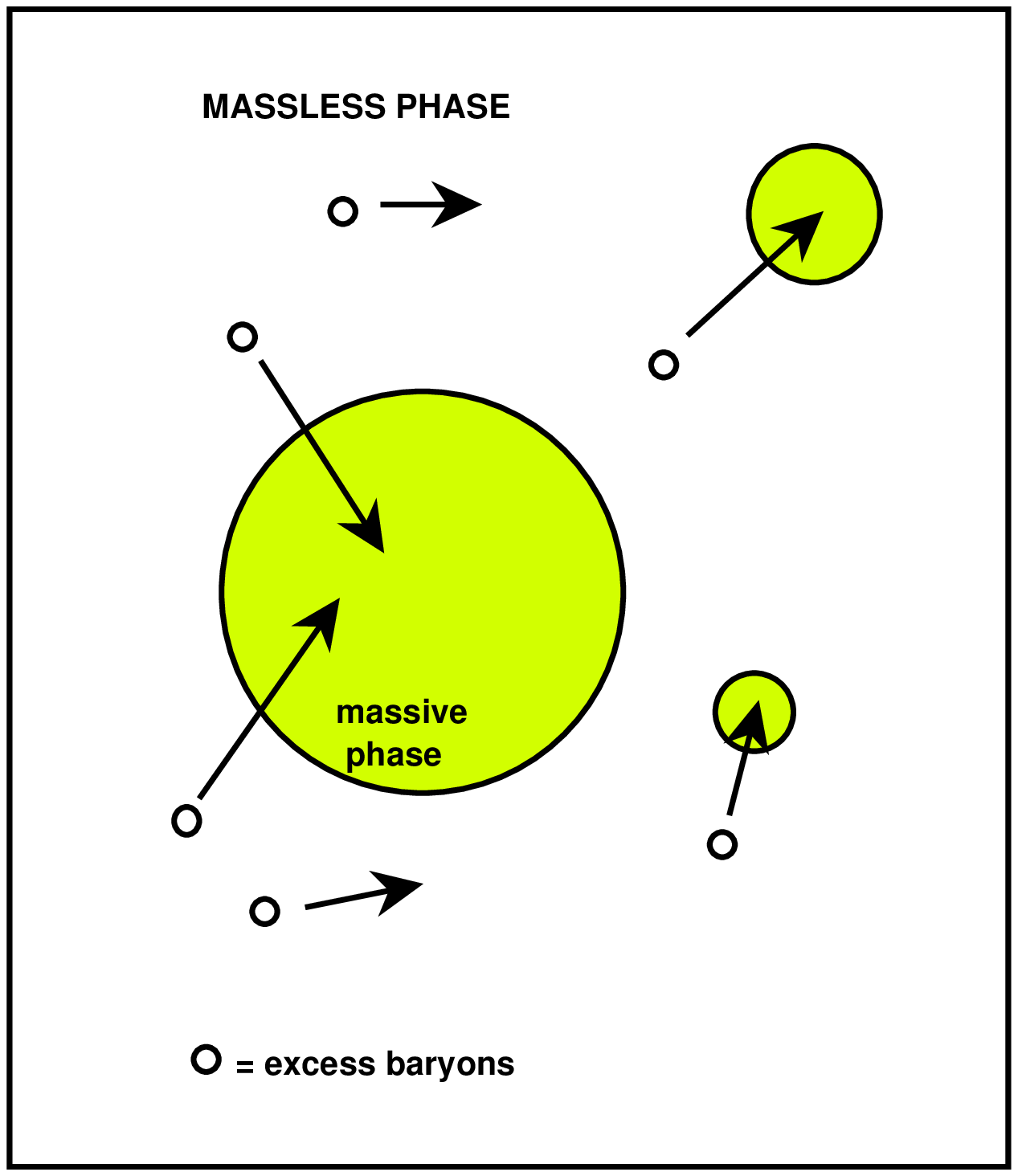,width=2.3in,height=2.3in}
\caption{Schematics of electroweak baryogenesis: In the symmetric massless
phase baryon number is violated, while in the massive broken symmetric phase
baryon number is conserved. The excess baryons generated in the massless 
phase will penetrate the bubbles, generating the excess baryon number in
the Universe.}
\end{figure}

There have been several different versions of electroweak baryogenesis
in the past five years or so, motivated mostly by the difficulty of
generating the right amount of CP violation in the Standard Model of
particle physics. These so-called extensions to the Standard Model come
in many different flavours, but are usually able to generate a much larger
amount of CP violation and thus of baryonic excess, even if at the cost of
introducing more arbitrary parameters. On the other hand, one could argue that
electroweak baryogenesis calls for physics beyond the Standard Model, another
beautiful illustration of the cosmology/particle physics interface implemented
through cosmological phase transitions. 

Although much progress has been made in our understanding of the dynamics of
cosmological phase transitions and their impact on the history of the Universe,
it should be clear that the future of this field is still quite open. As this
author has shown in a series of articles, the use of typical bubble nucleation
mechanisms to describe these transitions may be naive, the truth lying 
somewhere in between the two ``archetypes'' of continuous and discontinuous
phase transitions \cite{GLEISER,GLEISER2}. 
Since cosmological phase transitions are the
main link between micro and macro physics, we should expect many surprises 
in the years to come.

\end{document}